\documentstyle[aps,graphics,epsfig,prl,twocolumn]{revtex}
\begin{document}
\draft
\title{Anomalous Quasiparticle Lifetime in Graphite: Band Structure Effects}
\author{Catalin D. Spataru$^{1,2}$, Miguel A. Cazalilla$^{3}$,
 Angel Rubio$^{4,5}$, Lorin X. Benedict$^{6}$,\\Pedro M. Echenique$^{4,5,7}$,
and Steven G. Louie$^{1,2}$}
\address{$^{1}$Department of Physics, University of California
at Berkeley, Berkeley, CA 94720, USA\\
$^{2}$Materials Sciences Division, Lawrence Berkeley National 
Laboratory,
Berkeley, CA 94720, USA\\
$^{3}$Department of Physics, Brown University,
Providence, RI 02912-1843, USA\\
$^{4}$Materialen Fisika Saila,
Kimika Fakultatea, 1072 postakutxatila, 20080 Donostia, Spain\\
$^{5}$Donostia International Physics Center,
Paseo Manuel de Lardizabal 4,
20018 Donostia, Spain\\
$^{6}$H Division, Physics and Advanced Technologies Directorate, 
Lawrence Livermore National
Laboratory, University of California, Livermore, CA 94550,USA\\
$^{7}$Centro Mixto UPV-CSIC, 20080 Donostia, Spain}
\date{\today}
\maketitle

\begin{abstract}
We report {\it ab initio} calculation of quasiparticle 
lifetimes in graphite, as determined from the imaginary part of 
the self-energy operator within the $GW$ aproximation. The inverse
lifetime in the  energy range from $0.5$ to $3.5$ eV above the Fermi level
presents significant deviations from the quadratic behavior {\it naively}
expected from Fermi liquid theory. The deviations are 
explained in terms of the unique features of the band structure
of this material. We also discuss the experimental results from
different groups and make some predictions for future experiments.
\end{abstract}

\pacs{71.10.-w,79.60.-i,81.05.Uw}

  The Fermi liquid theory~\cite{AGD75} (FLT) has been one of the most successful
paradigms to describe the  metallic state of condensed matter.
According to it, the excitation spectrum in a metal can be understood 
in terms of
the quasiparticle (QP) concept: Quasiparticles are in one-to-one
correspondence with the one-electron excitations of a non-interacting
Fermi system, but their properties are modified by the interactions.
In a Fermi liquid, for energies very near the Fermi level,
quasiparticles have a finite lifetime ($\tau$), which is inversely proportional to the square of
the excitation energy\cite{AGD75,AshcroftMermin76}, i.e.
$\tau \propto (E-E_{F})^{-2}$. It is tempting to claim a breakdown of the paradigm when deviation from this behavior is found for a metallic system.
Recent experiments by Xu and coworkers~\cite{Xu96} on graphite have
reported an anomalous behavior of the QP lifetime
for $E-E_{F}$ in the range of $0.4$ to $2$ eV.
These authors observed that $\tau \propto (E-E_{F})^{-1}$.
In this letter, we argue that, when the band structure of
graphite is taken into account to compute the QP lifetime within
FLT, important deviations from $(E-E_{F})^{-2}$ behavior
are found in the experimentally accessible energy range.
We have found the QP lifetime to be strongly dependent on the
wave vector direction and that, when averaged over the first Brioullin
zone (BZ), the energy dependence of $\tau$ cannot be fitted to a simple power
law over the considered energy range, from $0.5$ to $2.5$ eV. All
these features find a simple explanation in terms of the dependence of the 
electron self-energy on the peculiarities of the band structure of graphite.

  In the experiments  carried out by Xu {\it et al.} the sample
(in this case, Highly Oriented Pyrolytic Graphite, HOPG) is excited
by a laser pulse and a second delayed pulse is used to probe the
population of the excited states. Varying the delay time
between the  first (pump) and the second (probe) pulses  provides
a powerful, though still poorly understood, spectroscopic technique known as
two-photon photoemission (2PPE). Subsequent measurements
by Ertel and coworkers~\cite{Ertel99},  using
very similar techniques, have partially confirmed
these results. They
found that $\tau \propto (E-E_{F})^{-\alpha}$ with
$\alpha\approx 1.2 - 1.3$  provided a better fit to their 2PPE data,
for $0.2$ eV $< E-E_{F} < 1.6$ eV. It should also be pointed out
that the absolute  value of $\tau$  seems to depend on  the method
chosen to fit the 2PPE signal as well as the way in which the
transport processes at the surface are accounted for~\cite{Ertel99}.
However, in both  experiments  the deviation from the $(E-E_{F})^{-2}$
dependence (found in many simple and noble 
metals~\cite{Campillo99,Dolado00}) seems clear.

  Graphite is a layered semimetal. The
coupling between layers (i.e., along the $c$-direction) is
much weaker than within the layers (i.e. in the $ab$-plane),
where strong bonds of covalent type keep the atoms arranged in a
honeycomb lattice with two atoms per unit cell. This arrangement makes
the valence and conduction $\pi$-like bands touch at the corners of the
hexagonal first BZ. Interlayer coupling in an ABAB... stacking sequence, however, enhances this overlap and shifts the Fermi level,  thus leading to
the formation of electron and hole pockets responsible for the
semimetallic character of graphite. Gonz\'alez
{\it et al.}~\cite{Gonzalez96}
have put forward  an explanation that relates the apparent linear
dependence of the inverse QP lifetime to
the structure of the  degenerate $\pi$-bands  near $E_{F}$
and  the weak coupling between layers. These two factors result
in a  small density of states at $E_{F}$. As consequence,
the screening length in graphite is very 
large~\cite{Gonzalez96,Screening}, a fact that
strongly affects QP lifetimes. However, the theory developed
by Gonz\'alez and coworkers
relies on a simplified  model for the band  structure of graphite,
which assumes that the $\pi$-bands disperse linearly with the
wave vector parallel to the $ab$-plane (${\bf k}_{||}$) up to an energy
of $\approx \pm 3$ eV around $E_{F}$.  Such a simplified band structure is
obtained as a low energy limit of a tight-binding fit of the
$\pi$-bands~\cite{Gonzalez94_99}, and completely neglects interlayer hopping.
Our and previous band-structure calculations, however,
indicate that, along some $\bf k$-directions, the dispersion is strongly
non-linear already  for $E-E_{F} \approx 1.5$ eV.  Moreover, hopping between
graphite layers, although smaller than intralayer hopping,
is  not negligible ($\approx 0.25$ eV) and leads to a splitting of the
$\pi$-bands at all energies.

   Motivated by these findings and the experiments mentioned
above,  we have evaluated the  QP lifetime accounting for the
band structure of graphite \cite{KohnSham65}. The lifetime for a QP with 
wave vector 
${\bf k}$ and band
index $n$ was obtained from  the imaginary part  of the electron
self-energy  $\Sigma(E)$~\cite{Campillo99,Echenique00}:
\begin{eqnarray*}
\frac{1}{\tau_{n\bf k}}=-2 \: \langle {n\bf k|}
{\rm Im}\: \Sigma (E_{n\bf k})|{n\bf k} \rangle,
\end{eqnarray*}
where we have neglected the renormalization of the QP wave function
and energy by taking the expectation value over the corresponding
LDA orbital $|n \bf k \rangle$ and setting $E= E_{n {\bf k}}$, the 
LDA eigenvalue.
$\Sigma(E)$ was obtained within the  $GW$
approximation~\cite{Hedin65_69,Hybertsen86}. Since we are interested in
a spectral property, namely the QP lifetime, it is sufficient to stop 
at this level~\cite{Holm98}.  Thus
the screened electron-electron interaction can be calculated
within the random phase approximation (RPA)~\cite{Pines63}
in terms of the dielectric matrix $\epsilon_{{\bf G}{\bf G'}}({\bf
q},\omega)$~\cite{Campillo99,Echenique00}. Finally, since the 2PPE experiments
can only probe the energy dependence in the relaxation time  of the excitations
generated by the pump pulse,  we have averaged  $\tau_{n\bf k}^{-1}$
over the states with energy  $E = E_{n {\bf k}}$ to obtain an
energy-dependent inverse lifetime, $\tau^{-1}(E)$~\cite{Campillo99}.
\begin{center}
\vspace{0.5cm}
\epsfig{file=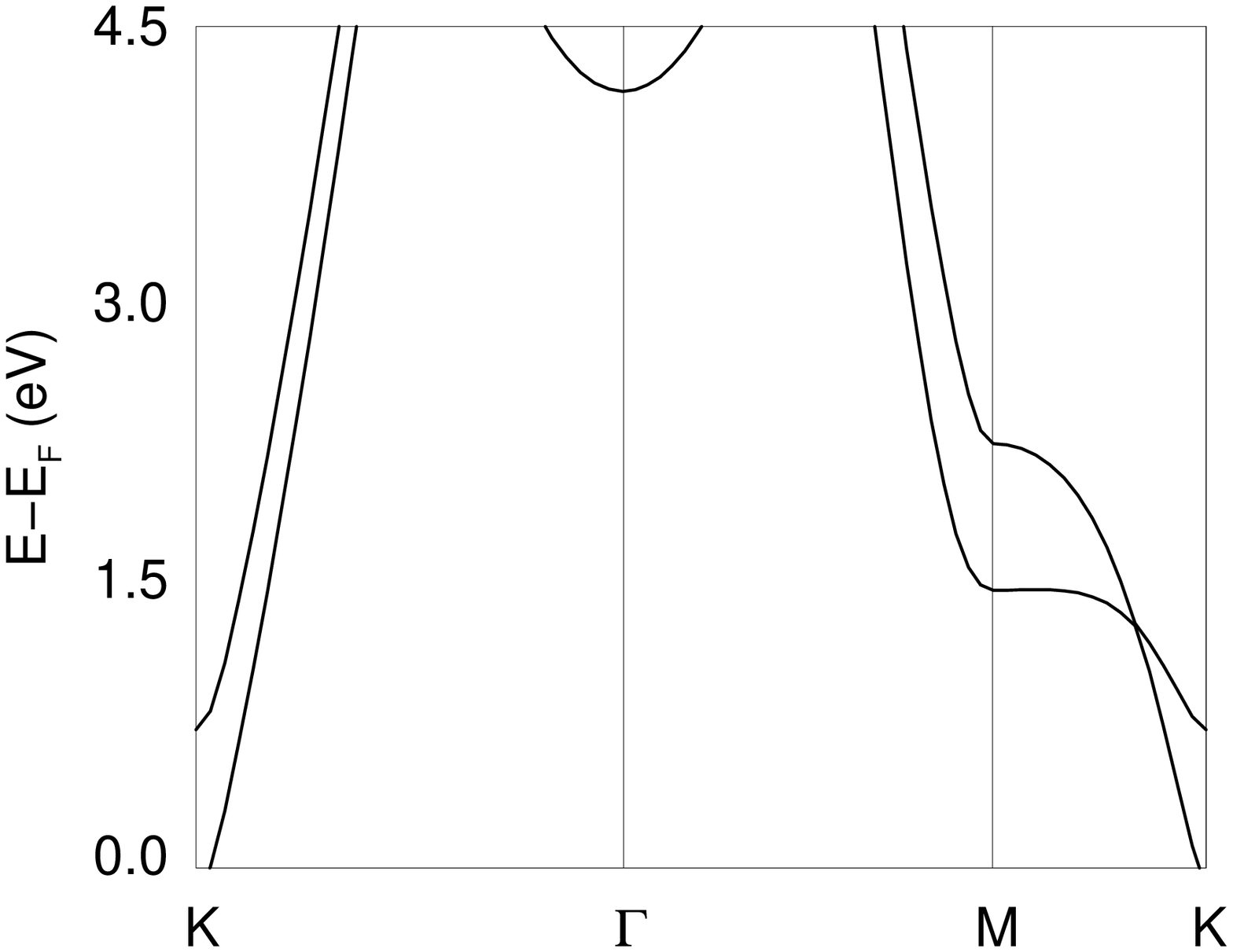, width=8.0cm}
\end{center}

\begin{figure}
\center
\caption{Graphite $\pi$-band structure near some relevant points $M$,
$L$ and $K$ in the energy range considered in this work.}\label{fig1}
\end{figure}
   The reason that the average QP lifetime cannot be fitted to
a simple power law over the considered energy range is because of the
smallness of
$\tau^{-1}$ in the neighborhood of the $M$ and $L$ points of the first
BZ, where the $\pi$-bands exhibit a saddle point (see Fig.1).
This band topology forces  a QP excited near these points  to
decay into a lower energy state by transferring a relatively large momentum (${\bf q}_{||}$)
parallel  to the $ab$-plane.
On the other  hand, the large  screening length referred to above favors small
${\bf q}_{||}$ electron-electron scattering and, at the same time,
the band structure below $1.5$ eV leads to a small density of
low energy excitations available at large momenta. This  point is
illustrated in  quantitative detail by Fig.2, where it can
be seen that the spectral weight of the energy loss function
in the low frequency region rapidly decreases with increasing $|{\bf q}_{||}|$.
This feature is also present
in ${\rm Im}\:\epsilon_{{\bf 0}{\bf 0}}({\bf q},\omega)$ (not
shown here for simplicity), which exhibits a similar structure,
revealing that the low  energy excitations are indeed electron-hole
pairs and not plasmons. Thus the main decay channel for QP's  is
provided by electron-hole pair creation
and not by plasmon excitation as speculated in Ref.~\cite{Xu96} (see, however,
Refs.~\cite{Gonzalez96,Comment}).
The coarseness of the grid used in
our calculation precludes us from obtaining the lifetime
below $0.5$ eV where we expect other decay channels such as
phonons or low energy plasmons~\cite{Jensen91} to be very important as
well.
\begin{center}
\vspace{0.5cm}
\epsfig{file=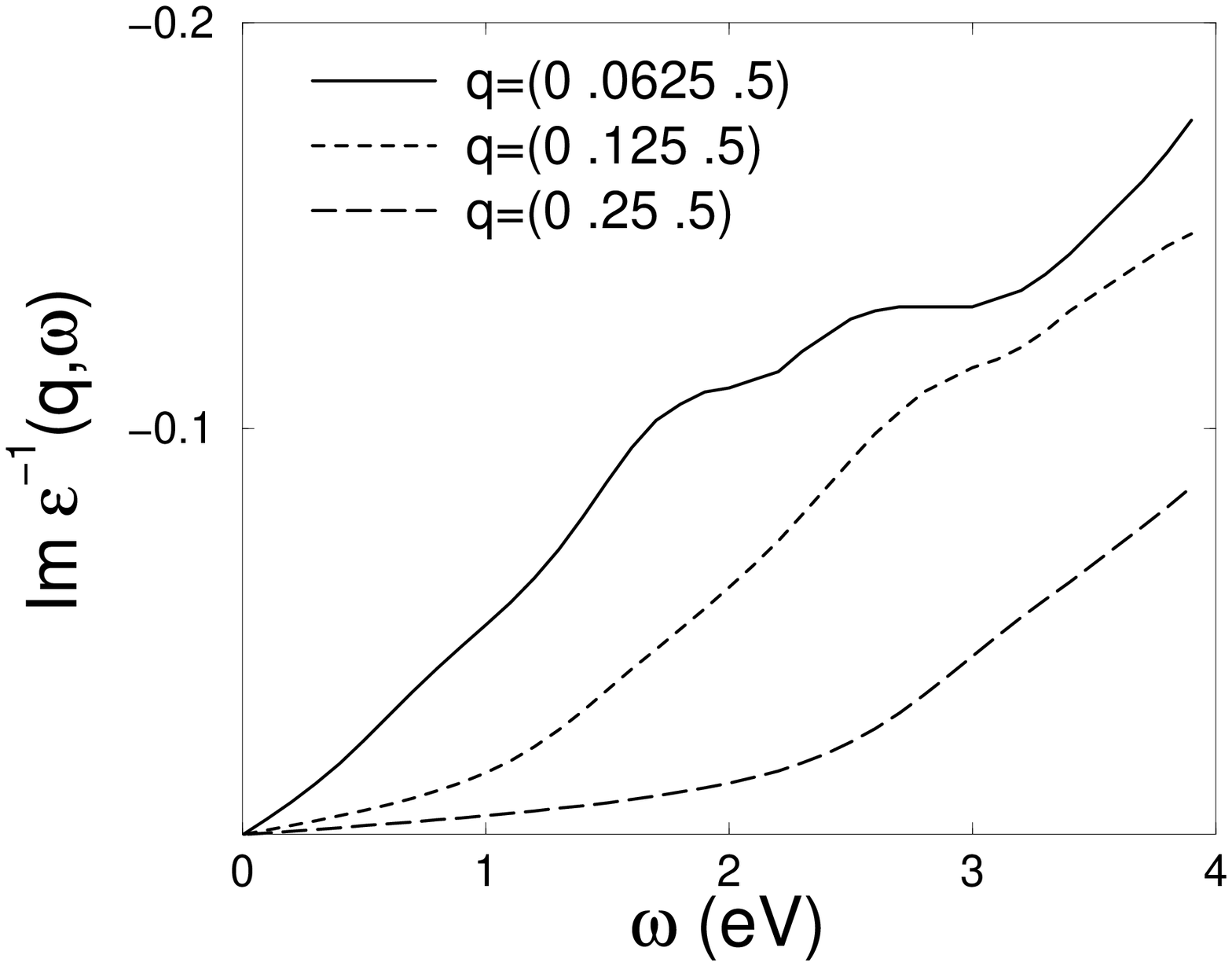, width=8.0cm}
\end{center}

\begin{figure}
\center
\caption{Energy loss function, ${\rm Im}\: \epsilon^{-1}_{\bf 0 0}({\bf
q},\omega)$ vs. frequency for different values of
$\bf q$.}\label{fig2}
\end{figure}

   In Fig.3 our results for $\tau^{-1}_{n{\bf k}}$ are plotted
vs. the LDA energies, $E_{n{\bf k}}$, for several ${\bf k}$-directions:
${K \to \Gamma}$, and $M \to \Gamma$.  Notice that the value of
the QP lifetime depends strongly on  ${\bf k}$. As discussed above, the 
unoccupied 
${\pi}$-bands possess a saddle  point near $M$ (cf. Fig.1), where 
the QP lifetime increases dramatically. This explains the rapid
drop of $\tau^{-1}_{n{\bf k}}$  along $M \to \Gamma$,
for $E-E_{F} \approx 1.5$ eV. Above this energy,
the states around $M$ become available for a QP to decay  and
$\tau^{-1}_{n{\bf k}}$ becomes appreciable. As $E-E_{F}$ increases further, 
more bands come into play
and thus the lifetime aquires two values near $E-E_{F}\approx
2.3$ eV, which correspond to states
with similar energies but different band index, $n$. The
splitting of the lowest energy bands along the $K \to \Gamma$
point also has the same effect.  This splitting is due to the
interlayer hopping, $t_{\perp} \approx 0.25$ eV, which lifts the
degeneracy of the $\pi$-bands near $E_{F}$.
\begin{center}
\vspace{0.5cm}
\epsfig{file=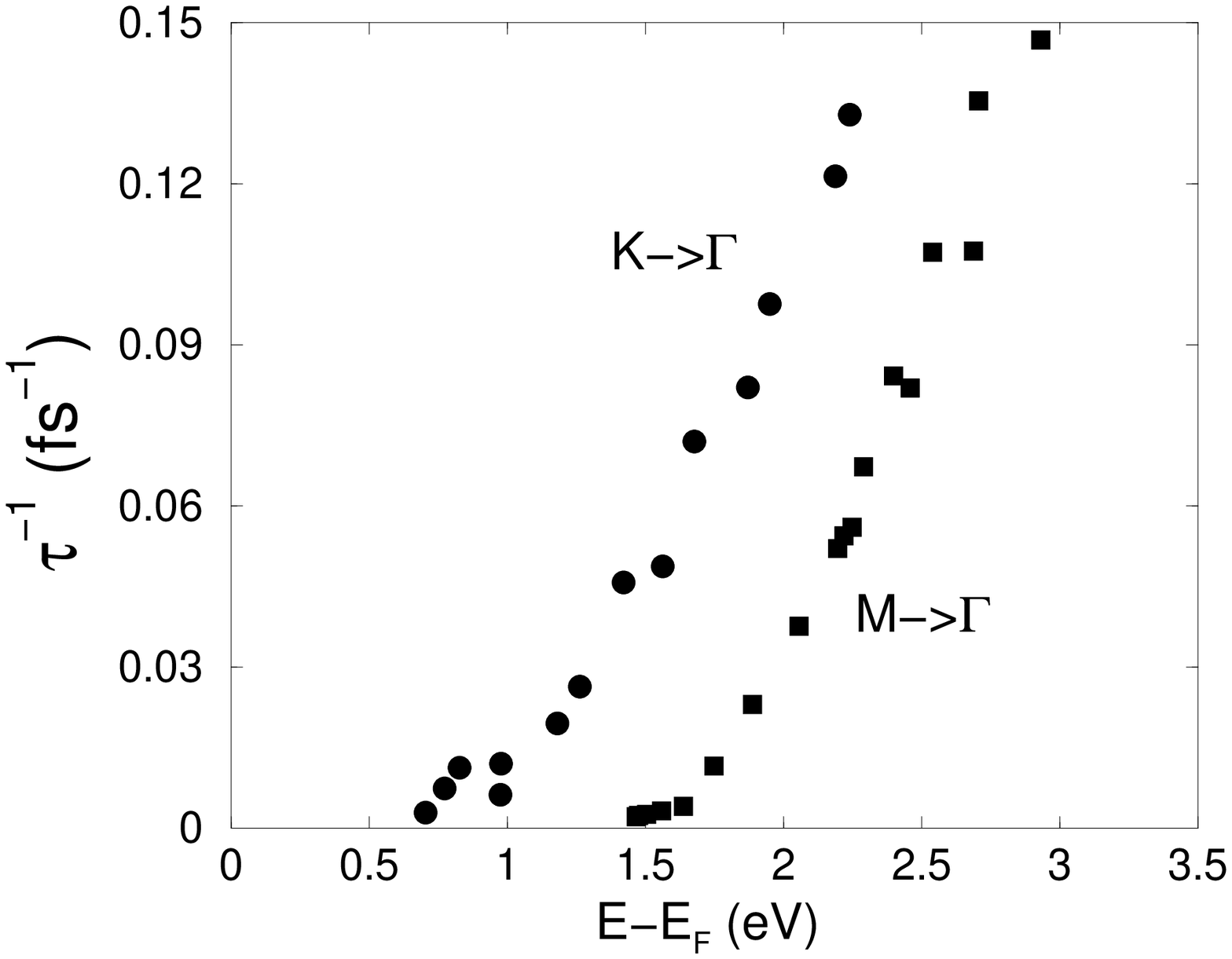, width=8.0cm}
\end{center}

\begin{figure}
\center
\caption{Inverse QP lifetime $\tau^{-1}_{n{\bf k}}$ vs. LDA eigenvalues, for two representative
directions $K \to \Gamma$ (circles) and $M \to \Gamma$ (squares).}\label{fig3}
\end{figure}
  As to the $K \to \Gamma$ direction,
one can regard the behavior of $\tau^{-1}_{n{\bf k}}$
as controlled by two parameters: The excitation energy,
$\epsilon =  E-E_{F} \approx v_{F} \: |{\bf k}_{||}|$, where
$v_{F}$ is the Fermi velocity, and the interlayer hopping
$t_{\perp}$. If we focus  on the average
behavior of the lifetime along this direction,  given the small
splitting of the bands, we can drop the band index $n$.
Let $\tau^{-1}(\epsilon, t_{\perp})$ be
the  QP inverse lifetime. On dimensional grounds,
$\tau^{-1}(\epsilon,t_{\perp}) = g^{2} \epsilon\,
f(t_{\perp}/\epsilon, g)$, where $g \propto e^{2}/v_{F}$ is the dimensionless
constant ``charge'' that characterizes the strength of the
electron-electron interaction~\cite{Gonzalez96,Gonzalez94_99} and $f(x,g)$
is a scaling function.
If $t_{\perp} = 0$ then the model of Ref.~\cite{Gonzalez96} would
apply for the QP states along  $K \to \Gamma$. Therefore, $\tau^{-1} \sim
g^{2}\: \epsilon$, which implies $f(0,g) =$ const. On the other
hand, for $t_{\perp}$ finite the QP
lifetime is given by Fermi liquid theory~\cite{Gonzalez94_99} so that 
$\tau^{-1} \sim g^{2}
\epsilon^{2}$ as $\epsilon \to 0$, and therefore $f(x \gg 1, g) \sim
1/x$. This means  that we can expect a  {\it crossover} from
a regime where  $\tau^{-1}$ has an 
approximate linear behavior with $\epsilon$
to the quadratic behavior expected from FLT.
The latter can be hardly seen in our results,
which only extend down to $\epsilon \approx 0.5$ eV,
but, since we rely on the QP picture of FLT for our calculations,
it should be recovered sufficiently  close to $E_{F}$.
On the other hand, for $\epsilon$ well above $t_{\perp}$, we
have approximately $\tau^{-1} \propto \epsilon$ along $K \to \Gamma$,
as seen in Fig.3.
Thus our results for this direction
agree qualitatively with the linear dependence found
by Gonzalez {\it et al.}~\cite{Gonzalez96} on the assumption that
$\epsilon \approx v_{\rm F}  |{\bf k}_{||}|$ but
  using a  different approach.

\begin{center}
\vspace{0.5cm}
\epsfig{file=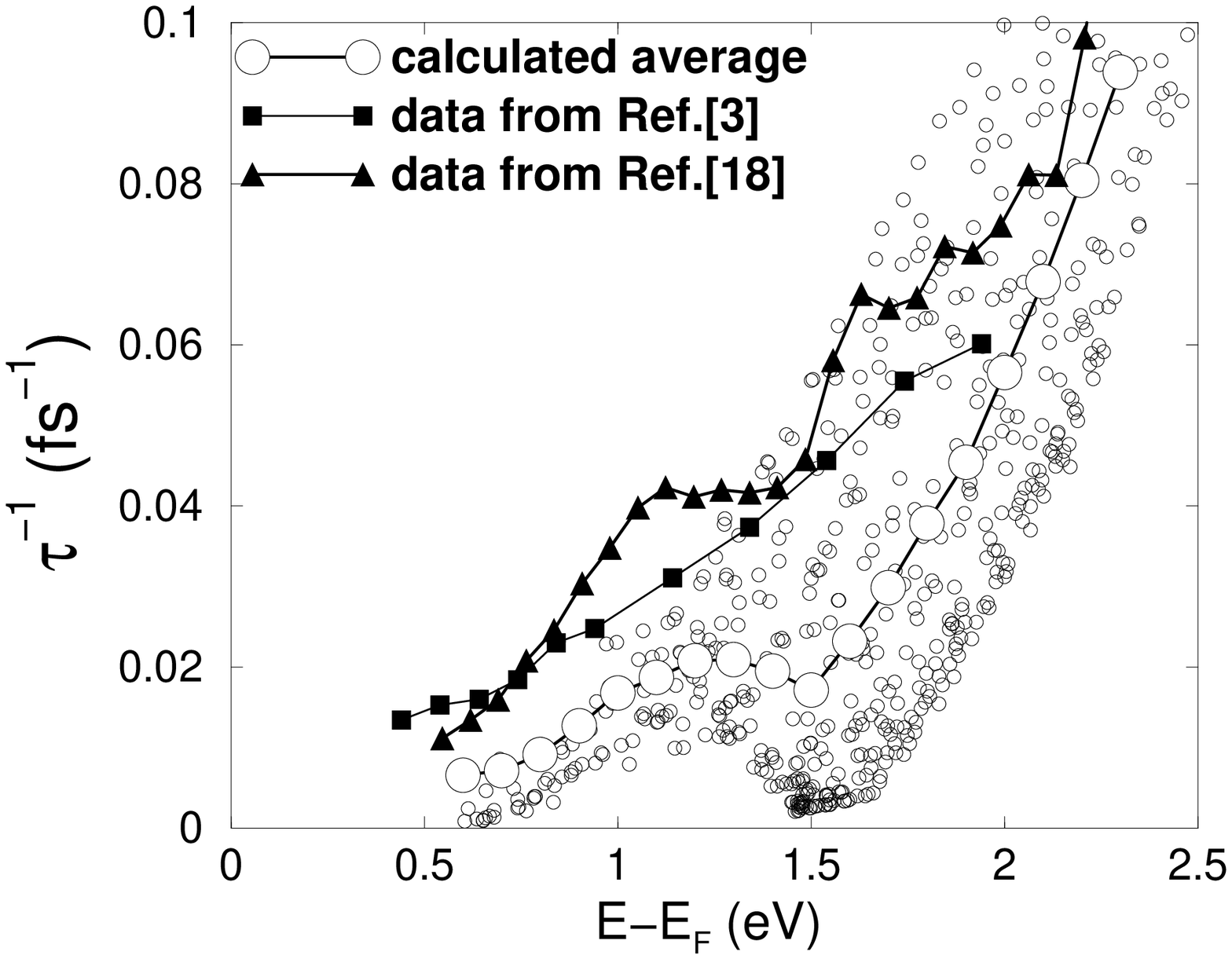, width=8.0cm}
\end{center}

\begin{figure}
\center
\caption{Average inverse QP lifetime $\tau^{-1}(E)$ vs. QP energy $E$: present calculation (circles) and experimental results from Ref. [3] (squares) and Ref. [18] (triangles).}\label{fig4}
\end{figure}

Finally, in Fig.4 we show the results for the inverse lifetime
of all the ${\bf k}$-points considered and the resulting average,
which shows three different regimes: $\tau^{-1}$ increases monotonically
between ${0.5}$ and ${1.0}$ eV. From there and up to ${1.5}$ eV, it remains
approximately constant and even decreases due to the long QP lifetime
found near $M$ and $L$ at $E-E_{F}\approx 1.5$ eV, which was explained
above. Above $1.5$ eV $\tau^{-1}$ increases rapidly and, 
as the energy increases, band structure effects should  become less and less important.
Also, for the sake of comparison we have plotted the experimental
results reported in Ref.~\cite{Xu96}, and new experimental results of Moos 
et al. \cite{HertelUU}. A word of caution is in order
as the precise interpretation of what is measured in these experiments
in terms of QP lifetime is far from clear~\cite{Ertel99,Campillo99,HertelUU};
results from different groups using very similar techniques and interpretations
do not always coincide. However, it is worth commenting on the
fact that the results by Xu and coworkers do not show any dip around $1.5$ eV that is predicted by our results but instead increase
monotonically in the whole energy range. At first, one could
blame the approximations made in the present calculation. Certainly,
if the renormalization of the QP energy is not neglected it can be
expected that the position of the saddle points near $L$ and $M$ may be slightly shifted upwards in energy and this would extend the range over which
$\tau^{-1}(E)$ increases monotonically with $E$. Also, detailed photoemission 
matrix elements, not considered here, may affect the magnitudes of the decay 
rates. Other QP decay
mechanisms such as electron-phonon coupling should be considered
as well. However, the results of Moos et al.~\cite{HertelUU} on pristine 
graphite surfaces show a dip in $\tau^{-1}$ around $1.5$ eV as predicted by the
calculations reported here. Furthermore,  they also observe a
strong dependence on the sample preparation, which might explain
discrepancies with the results from other groups.

   In conclusion, we have presented the results of a calculation of the
lifetime, $\tau$, which takes into account the band structure of graphite. 
We find important
deviations from the $\tau \propto (E-E_{F})^{-2}$ naively expected
from Fermi liquid theory. The deviations, however, find a
natural explanation given the peculiarities of the band structure of
graphite. The saddle point near the $M$ point leads
to a very large $\tau$ (alternatively a very small $\tau^{-1}$),
whereas the linearly dispersing bands near along the $K \to \Gamma$
direction lead to $\tau \propto (E-E_{F})^{-1}$ for energies well
above the interlayer hopping $\approx 0.25$ eV, in agreement with
previous work~\cite{Gonzalez96}. Finally, when
averaged over the Brioullin zone, we predict that $\tau^{-1}(E)$
cannot be fitted by a simple power law as it exhibits a
dip near $1.5$ eV. This behavior has been observed in
recent experiments~\cite{HertelUU}.

This work was supported by the NSF under Grant No. DMR0087088, 
Office of Energy Research, Office of  Basic Energy Sciences, Materials
Sciences  Division of the U.S. Department  of Energy under  Contract
No. DE-AC03-76SF00098, Basque Country University, Basque Hezkuntza Saila,
Iberdrola S.A. and DGES. C.~D.~S. would like to thank S. Ismail-Beigi for 
fruitful discussions. M.~A.~C. has been supported by the {\it Hezkuntza, 
Unibertstitate eta Ikerketa Saila} of the government of Basque Country
and NSF. Grant. no. DMR9712391. He 
is also grateful for the kind hospitality of the Donostia International 
Physics Center (DIPC), and for a number of useful discussions with F.~Guinea. 
Portions of this work were performed under the auspices of the U.S. Department of Energy by University of California Lawrence Livermore National Laboratory under contract No. W-7405-Eng-48. Computer time was provided by the DOE at the 
Lawrence Berkeley National Laboratory's NERSC center.


\begin{references}
\bibitem{Screening}
The screening length is inversely proportional to the density
of states at $E_{F}$. See, e.g., ~\cite{AshcroftMermin76} pag. 342.
\end{references}
\end{document}